\documentstyle[11pt,newpasp,twoside,epsf]{article}
\def\edcomment#1{\iffalse\marginpar{\raggedright\sl#1\/}\else\relax\fi}
\reversemarginpar

\begin{document}
\title{A solution to the transition phase in classical novae}
 \author{Alon Retter}
\affil{University of Sydney, School of Physics, University of Sydney, 
NSW, 2006, Australia}

\begin{abstract}
One century after the discovery of quasi-periodic oscillations in the 
optical light curve of Nova GK Per 1901 the cause of the transition 
phase in a certain part of the nova population is still unknown. Three 
years ago we suggested a solution for this problem and proposed a possible 
connection between the transition phase and intermediate polars (IPs). 
About 10\% of the cataclysmic variable population are classified as 
IPs, which is consistent with the rarity ($\sim$15\%) of the transition 
phase in novae. Recent observations of three novae seem to support our 
prediction. The connection is explained as follows: The nova outburst 
disrupts the accretion disc only in IPs. The recovery of the disc, a few 
weeks-months after the eruption, causes strong winds that block the 
radiation from the white dwarf, thus dust is not destroyed. If the winds 
are very strong as is probably the case in DQ Her (perhaps since its spin 
period is very short) this leads to a dust minimum.
\end{abstract}

\section{Introduction}

The optical light curve of a classical nova is typically characterized by  
a smooth decline. Certain novae show, however, a DQ Her-like deep minimum  
in the light curve while others have slow oscillations, during the so  
called `transition phase' (Fig. 1). The minimum is understood by the  
formation of a dust envelope around the binary system. It is still not  
known, however, what causes the oscillations during the transition phase  
(Warner 1995) and why only a small fraction (about 15\%) of the nova  
population has the transition phase.

\begin{figure}  
\plotone{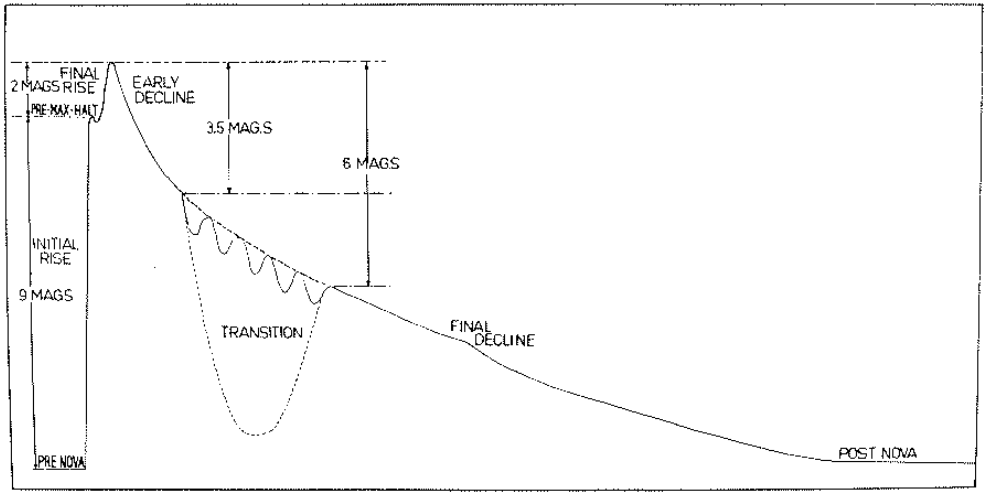}
\caption{A schematic diagram of the optical light curve of classical novae 
(adapted from Warner 1995)}
\end{figure}

The models offered so far to the oscillations during the transition phase 
(Bode \& Evans 1989; Leibowitz 1993; Warner 1995) are: 
 
\begin{itemize} 
\item 
  Oscillations of the common envelope that surrounds the binary system after  
the nova outburst. 
\item 
  Dwarf-nova outbursts. 
\item 
  Formation of dust blobs that move in and out of line of sight to the nova. 
\item 
  Oscillations in the wind (see also Shaviv, these proceedings). 
\item 
  Stellar oscillations of the hot white dwarf. 
\end{itemize} 
 
The first model can be rejected very easily as the common envelope phase 
(`fireball') lasts less than 1-2 days (Hauschildt, personal communication),  
so it is much shorter than the typical time scale of the transition phase.  
The second idea is almost certainly wrong as the accretion discs in  
post-novae are thermally stable (e.g. Retter \& Naylor 2000), so they  
cannot have dwarf-nova outbursts for at least many decades following the  
nova eruption. 
 
Retter, Liller, \& Gerradd (2000a) suggested another solution for this  
problem and we show here that it can be combined with two of the models 
listed above. The observations of Nova LZ Mus 1998, which had oscillations  
during the transition phase revealed a few periodicities in its optical  
light curve. Retter et al. thus classified LZ Mus as an IP candidate. IPs  
are cataclysmic variables in which the primary white dwarf has a moderate  
magnetic field and thus spins around its axis with a period shorter than  
the orbital period. Hernandz \& Sala (this volume) detected X-ray emission  
from LZ Mus three years after its outburst despite its extremely large  
distance. This fact is consistent with the IP classification. 
 
Retter et al. further proposed a possible connection between the transition  
phase and IPs and predicted that novae that have a transition phase should  
be IPs. About 10\% of the cataclysmic variable population are IPs, which is  
consistent with the rarity ($\sim$15\%) of the transition phase in novae.  
We note that these numbers represent lower limits as not all systems have  
been well studied. The correct ratios may be as high as $\sim$30\%.  
Recent observations of three young novae seem to support our idea. 
 
\section{The early presence of the disc in young novae} 
 
It was believed that the accretion disc is destroyed by the nova event  
and that it takes only a few decades for the disc to re-establish.  
In my Ph.D. thesis I studied this claim. Contrary to the common belief we  
found very strong evidence in at least two cases for the presence of the  
accretion disc only a few months after the nova outburst (Retter, Leibowitz,  
\& Ofek 1997; Retter, Leibowitz, \& Kovo-Kariti 1998). We note that the 
time-scales of the transition phase are similar to the time-scales of the 
appearance of the accretion disc in post-novae. It is still unknown 
whether the disc can survive the nova explosion. The various bright  
sources, which contribute to the optical light curve of post-novae, may,  
however, overcast the light from the disc and complicate the observations.  
We may have found now the solution to this question. This work suggests  
that the accretion disc is destroyed by the nova outburst only in IPs and  
that discs in other subclasses of CVs survive the nova event. 
 
\section{Observations and Analysis} 
 
Recent observations of three young novae seem to support the suggested 
connection between the transition phase and IPs. The optical light curve  
of Nova V1494 Aql 1999\#2 showed transition phase oscillations with a  
quasi-period of $\sim$7 days (Kiss \& Thomson 2000). A period of 3.2 h  
(presumably the orbital period) was discovered in its optical light curve  
(Retter et al. 2000b). Krautter et al. (2001) and Drake et al. (2002, in  
preparations) detected a 2523-s periodicity in two X-ray runs using  
Chandra. The short period can be interpreted as the spin period of the  
binary system and the nova can thus be classified as an IP. We note,  
however, the suggestion of an alternative model to the variation --  
stellar oscillations of the white dwarf (Krautter et al., these  
proceedings). 
 
On the other hand, Nova V382 Vel 1999 had a smooth decline in the optical 
(Liller \& Jones 2000) and X-ray observations did not reveal any  
short-term periodicity (Orio et al. 2001). So, it is unlikely that it is  
an IP. 
 
In addition, Nova V1039 Cen 2001 had oscillations during the transition phase 
as well (http://vsnet.kusastro.kyoto-u.ac.jp/vsnet/etc/drawobs.cgi?text=CENnova2001). Preliminary analysis of 6 nights of optical photometry using small  
telescopes + CCDs shows several peaks in its power spectrum (Fig. 2). They  
can be explained within the IP model, so this nova could also be an IP,  
consistent with our prediction. 
 
Additional possible support for our idea might come from the detection of  
oscillations in an X-ray source, which is located near the nucleus of M31.  
Osborne et al. (2001) proposed that the $\sim$865-sec periodicity represent  
the spin period of a magnetic white dwarf. The high X-ray luminosity and  
the transient nature of the object led them to propose that this is a recent  
novae. King, Osborne, \& Schenker (2002), however, argued against this  
suggestion. 
 
\begin{figure}  
\plotone{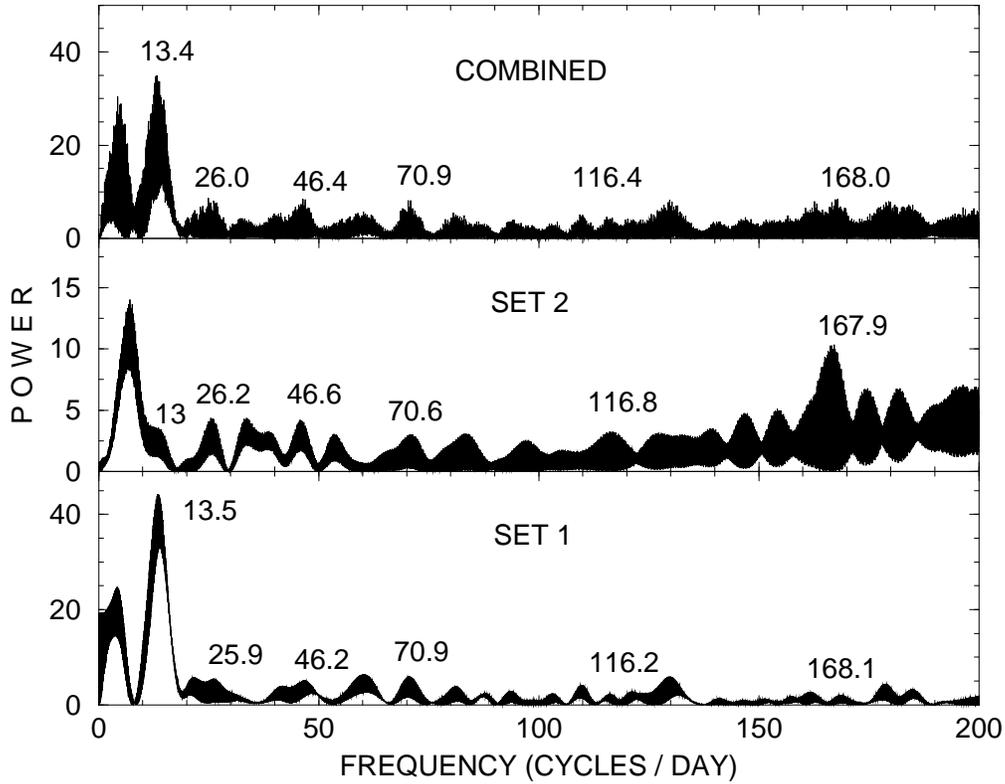}
\caption{Power spectra of six nights of photometry of Nova V1039 Cen 2001 
obtained in 2002. The lower two panels show two different sets and the top  
panel displays the combined data. The several groups of peaks may correspond  
to the orbital period, the spin period and some combinations of the two.  
V1039 Cen may thus be grouped with IPs.)}
\end{figure}

\section{Discussion} 
 
There seems to be a strong connection between the transition phase in  
classical novae and IPs. Two famous cases are Nova GK Per 1901  
(oscillations) and Nova DQ Her 1934 (minimum), which are well-known IPs.  
Another possible example is Nova V603 Aql 1918 (oscillations), which was  
suggested several times as an IP candidate (e.g. Schwarzenberg-Czerny,  
Udalski \& Monier 1992) and whose X-ray light curve show strong variations  
(Mukai et al., this volume). We note, however, that there has been a long  
argument whether this nova is indeed an IP (e.g. Mukai et al., these  
proceedings). 

We suggest that this link is connected with the presence of the accretion  
disc in post-novae. In IPs, the magnetic field truncates the inner part  
of the disc, making it less massive than in non-magnetic systems (this is,  
by the way, the reason why IPs do not have dwarf-nova outbursts or only  
have short and weak outbursts). The nova outburst can, therefore, disrupt  
the disc in IPs. Its re-establishment and the subsequent interaction with  
the magnetosphere of the primary white dwarf is a violent process that  
forms strong winds at the inner part of the disc. The winds block the  
radiation from the hot white dwarf and the dust is not destroyed. The  
accretion disc oscillates until finally reaching stability at the end of  
the transition phase. In non-magnetic systems the disc is barely disturbed  
and becomes stable much faster and in polars (AM Her systems) there is  
no accretion disc at all. In both groups there is no transition phase  
since dust cannot be formed.  
 
The long-term oscillations during the transition phase could be explained  
if the inner accretion disc lies very close to the co-rotation radius, 
so the disc and the magnetic field would rotate (almost) together at that  
point. The relative rotational timescale (the beat period) could be very  
long -- of the order weeks, as observed in the transition phase.  
 
If the winds are very strong, as might be the case in DQ Her (perhaps  
since its spin period is very short) this leads to the formation of more  
dust and to a minimum in the light curve. IPs with longer spin periods  
should show oscillations. Naturally, our model has a strong dependence on  
the inclination angle.  
 
Further X-ray and optical observations of novae that have a transition  
phase should confirm or refute our suggestion. 
 
\acknowledgments
 
We thank Graham Wynn for a very useful discussion on the cause of the 
long-term oscillations and Marina Orio for pointing to us Osborne et 
al.'s work. AR is supported by an ARC grant.

\end{document}